# Feature selection for transient stability assessment based on kernelized fuzzy rough sets and memetic algorithm

Xueping Gu [a,*], Yang Li [b,*], Jinghua Jia [c]

[a] *School of Electrical and Electronic Engineering, North China Electric Power University, Baoding 071003, China*

[b] *School of Electrical Engineering, Northeast Dianli University, Jilin 132012, China*

[c] *Hebei Power Dispatch and Communication Center, Shijiazhuang 050021, China*

**ABSTRACT:** A new feature selection method based on kernelized fuzzy rough sets (KFRS) and the memetic algorithm (MA) is proposed for transient stability assessment of power systems. Considering the possible real-time information provided by wide-area measurement systems, a group of system-level classification features are extracted from the power system operation parameters to build the original feature set. By defining a KFRS-based generalized classification function as the separability criterion, the memetic algorithm based on binary differential evolution (BDE) and Tabu search (TS) is employed to obtain the optimal feature subsets with the maximized classification capability. The proposed method may avoid the information loss caused by the feature discretization process of the rough-set based attribute selection, and comprehensively utilize the advantages of BDE and TS to improve the solution quality and search efficiency. The effectiveness of the proposed method is validated by the application results on the New England 39-bus power system and the southern power system of Hebei province.

## 1. Introduction

Transient stability is concerned with the ability of a power system to maintain synchronism when subjected to a severe disturbance, such as a short circuit on a transmission line [1]. Transient stability assessment (TSA) has been recognized as an important issue to ensure the secure and economical operation of power systems [2]. TSA may serve to check the operation mode arrangement of a power system as a beforehand analysis tool in a dynamic security assessment framework, and to trigger the emergency controls as a real-time stability prediction tool after faults. The both applications can effectively reduce the possibility and amount of load loss from transient instability and then improve the operational security and economical efficiency of the power system. Problems arising from introduction of new power market designs and growing presence of intermittent renewable power generation are nudging power systems toward potential dynamic instability scenarios. The traditional methods for transient stability analysis, such as time-domain simulation methods [2], transient energy function methods [3] and the extended equal-area criterion [4], can not well meet the requirements of online TSA for modern complex power systems. With the rapid development of computational intelligence such as decision trees (DT), artificial neural networks (ANN), and support vector machines (SVM), the pattern recognition-based TSA (PRTSA) methods have shown much potential for on-line application to power systems [5-13].

In the previous work of PRTSA [5, 7, 8, 12], much attention has been given to design of classifiers and their parameter tuning, and relatively less attention to the feature selection issue. From the pattern recognition principles, it is well-known that the excessive input features will induce heavy computational burden, reduce the accuracy of training models and even lead to the "curse of dimensionality" [14]. Meanwhile, for transient stability classification, the classification accuracy is in fact determined by separability of the input space created by the selected features [15]. Therefore, the study of feature selection is an issue with paramount importance for PRTSA.

Some useful explorations have been carried out on the feature selection of PRTSA [14-17]. Fisher's linear discriminant function is used to select neural network training features for power system security assessment in [14], but the effectiveness of the proposed method can not be theoretically guaranteed. In [15], a separability index as the classification criterion is defined through finding the 'inconsistent cases' in the sample set, and the breadth-first

---

*Corresponding author.

E-mail address: xpgu@ncepu.edu.cn (Xueping Gu), liyangnedu@gmail.com (Yang Li).

searching technique is employed to find the minimal or optimal subsets of the initial feature set as the ANN input. In [16], three dimensionality reduction methods, including sensitivity index, sensitivity analysis and principal component analysis, are used to reduce the input space dimension for ANN-based TSA. However, both the proposed methods in [15] and [16] are only tested on a small test power system, and their effectiveness on practical complex power systems needs to be further validated because of the heavy computation burden. In [17], correlation analysis and principle component analysis are used as feature reduction techniques to reduce the number of the input features, but the used original input features are single-machine features, rather than system-level features, which are unsuitable for stability analysis of large-scale power systems. In feature selection, there are two key problems: feature evaluation metric and search strategies. The rough set theory (RST)-based separability criterion can be used as an effective feature evaluation index [18]. However, as most datasets contain real-valued features, it is necessary to perform feature discretization beforehand when using RST, which will inevitably cause quantization error and information loss problem [19]. As to the optimal feature subsets, the current search strategies, such as the sequence of feature selection techniques, TS and breadth-first search methods, have disadvantages of low efficiency and/or local optimum trapping.

The kernelized fuzzy rough set (KFRS) is an effective tool in dealing with uncertainty in data analysis [20], which combines the advantages of both kernel methods and RST. The Memetic algorithm (MA) is a stochastic optimization algorithm based on the imitation of cultural evolution [21], which has been successfully applied to solve many complex optimization problems [22, 23]. In this paper, the MA is combined with the KFRS to be used for feature selection of PRTSA.

In recent years, wide-area measurement systems (WAMS) make it possible to obtain the synchronized real-time state information, and this brings new ideas and opportunities to transient stability assessment and prediction [24-26].

In view of the current status of the PRTSA feature selection, a new feature selection method based on KFRS and MA is proposed for real-time transient stability prediction in this paper. Considering the possible real-time information provided by WAMS, a group of system-level classification features are extracted from the power system operation parameters to build the original feature set. By defining a KFRS-based generalized classification function as the separability criterion, the memetic algorithm based on binary differential evolution (BDE) and Tabu search (TS) is then employed to obtain the optimal feature subsets with the maximized classification capability. The proposed method is verified by the numerical results on the New England 39-bus power system and the southern power system of Hebei province.

## 2. KFRS and the class separability criterion

### 2.1. KFRS

The main idea of KFRS is as follows: kernel functions are employed to compute the fuzzy *T*-equivalence relations between samples, thus generating fuzzy information granules in the approximation space; subsequently fuzzy granules are used to approximate the class demarcation based on the concepts of fuzzy lower and upper approximations, and build a kernelized model of fuzzy rough sets [20].

A classification task can typically be formulated as $<U, A, D>$, where $U$ is the nonempty and finite set of samples, $A$ is the set of features characterizing the classification, $D$ is the class attribute which divides the samples into subset $\{d_1, d_2, \cdots, d_m\}$.

Given an arbitrary subset of features $B \subseteq A$ and $B \neq \varnothing$, a fuzzy *T*-equivalence relation $R$ over $U$ can be generated, where $\forall x, y, z \in U$, $R(x,x) = 1$; $R(x,y) = R(y,x)$ and $T(R(x,y), R(y,z)) \leq R(x,z)$, $T$ is a triangular norm. The fuzzy information granules induced by relation $R$ and $x_i$, denoted by $FIG_R(x_i)$, is defined as

$$FIG_R(x_i) = r_{1i}/x_1 + r_{2i}/x_2 + \cdots r_{ji}/x_j + \cdots + r_{ni}/x_n \tag{1}$$

where $r_{ji}$ is the similarity degree of samples $x_i$ and $x_j$. According to the definitions of lower and upper approximations, the memberships of a sample $x$ to lower and upper approximations of the class $d_i$ are computed by

$$\begin{cases} \underline{R_S}d_i(x) = \inf_{y \in U} S(1 - R(x,y), d_i(y)) \\ \overline{R_T}d_i(x) = \sup_{y \in U} T(R(x,y), d_i(y)) \end{cases} \quad \text{or} \quad (2)$$

$$\begin{cases} \underline{R_\theta}d_i(x) = \inf_{y \in U} \theta(R(x,y), d_i(y)) \\ \overline{R_\sigma}d_i(x) = \sup_{y \in U} \sigma(N(R(x,y)), d_i(y)) \end{cases} \quad (3)$$

where $\underline{R_S}d_i(x)$ and $\underline{R_\theta}d_i(x)$ are the degrees of certainty of the sample $x$ belonging to decision $d_i$, whilst $\overline{R_T}d_i(x)$ and $\overline{R_\sigma}d_i(x)$ are the degrees of possibility of the sample $x$ belonging to decision $d_i$.

In Theorem 1, Moser showed that part of kernel functions can be introduced to get fuzzy $T$-equivalence relations.

**Definition 1. [27]** *Give a nonempty and finite set $U$, a real-valued function $k: U \times U \to R$ is said to be a kernel if it is symmetric, that is, $k(x,y) = k(y,x)$ for all $\forall x, y \in U$, and positive-semidefinite.*

**Theorem 1. [28]** *Any kernel $k: U \times U \to [0,1]$ with $k(x,x) = 1$ is (at least) $T_{\cos}$-transitive, where $T_{\cos}(a,b) = \max(ab - \sqrt{1-a^2}\sqrt{1-b^2}, 0)$.*

Obviously, the Gaussian kernel $k(x,y) = \exp(-\|x-y\|^2/\delta)$ satisfies the above conditions, where $\delta$ is the width of the Gaussian. Therefore the relations computed with Gaussian kernel are fuzzy $T$-equivalence relations between samples. Then the formulae for computing the memberships of lower and upper approximations can be obtained by

$$\begin{cases} \underline{k_S}d_i(x) = \inf_{y \notin d_i}(1 - k(x,y)) \\ \underline{k_\theta}d_i(x) = \inf_{y \notin d_i}(\sqrt{1 - k^2(x,y)}) \\ \overline{k_T}d_i(x) = \sup_{y \in d_i} k(x,y) \\ \overline{k_\sigma}d_i(x) = \sup_{y \in d_i}(1 - \sqrt{1 - k^2(x,y)}) \end{cases} \quad (4)$$

$\underline{k_S}d_i(x)$ and $\underline{k_\theta}d_i(x)$ are the degrees the sample $x$ certainly belongs to class $d_i$, while $\overline{k_T}d_i(x)$ and $\overline{k_\sigma}d_i(x)$ are the degrees this sample $x$ possibly belongs to class $d_i$.

### 2.2. Class separability criterion

In order to enhance the robustness of classification index, $N_k$ nearest neighbors of each sample from each different class and from the same class are comprehensively considered. Given $<U, A, D>$, a KFRS-based generalized classification function $gc(D)$ is used as the class separability criterion in this paper.

$$gc(D) = \left[ g\gamma_B^\theta(D) + g\omega_B^{\theta-\sigma}(D) \right]/2 \quad (5)$$

where, $g\gamma_B^\theta(D)$ and $g\omega_B^{\theta-\sigma}(D)$ are respectively the generalized dependency function and generalized classification certainty function with $N_k = 3$, as given by:

$$g\gamma_B^\theta(D) = \frac{1}{(I-1)N_k|U|} \sum_{x_i \in U} \sum_{d \in D, d \neq d_i} \sum_{y \in H_d^i} \sqrt{1 - k(x,y)^2}$$

$$g\omega_B^{\theta-\sigma}(D) = \frac{1}{(I-1)N_k|U|} \sum_{x_i} \left\{ \sum_{H^i} \sqrt{1 - k(x_i, H_j^i)^2} - \sum_l \sum_m \left[ 1 - \sqrt{1 - k(x_i, M_{lm}^i)^2} \right] \right\},$$

where $I$ is the number of classes, $U$ is divided into $D = \{d_1, d_2, \cdots, d_I\}$ with the decision attribute, the feature space $B \subseteq A$, $H_{d_i}^i$ denotes the nearest $N_k$ samples of $x_i$ from each class except $d_i$ ($d_i$ is the decision of $x_i$), $H^i$ denotes the nearest $N_k$ samples of $x_i$ from the same class and the nearest $N_k$ samples of $x_i$ from each different classes, denoted by

$$M^i = \begin{vmatrix} M^i_{11} & M^i_{12} & \cdots & M^i_{1N_k} \\ M^i_{21} & M^i_{22} & \cdots & M^i_{2N_k} \\ \vdots & \vdots & \ddots & \vdots \\ M^i_{(I-1)1} & M^i_{(I-1)2} & \cdots & M^i_{(I-1)N_k} \end{vmatrix}, \quad M^i_{lm} \text{ is the } m\text{th nearest sample of } x_i \text{ from class } d_l.$$

## 3. Feature selection based on MA

In the mathematical model for feature selection built in this paper, a KFRS-based generalized classification function is employed as the separability criterion and the candidate feature subsets are generated by using a memetic algorithm based on BDE and TS to find the feature subset with maximized separability. BDE is used to global search for the entire solution space, and the neighborhood of elite solutions is searched by TS. Therefore, the optimization process is transformed into a dynamic evolution problem of intelligent systems.

### 3.1. Mathematical model of feature selection

Given a feature set $F_{all} = \{f_1, f_2, \cdots, f_i, \cdots, f_N\}$, a subset $F_s$ of which can be determined by $S = \{s_1, s_2, \cdots, s_i, \cdots s_N\}$, $s_i \in \{0,1\}$, $i = 1, 2, \cdots, N$. Here, $s_i$ denotes whether the $i$th feature $f_i$ is selected. If $f_i$ is selected, then $s_i = 1$; otherwise, $s_i = 0$. The classification performance of $F_s$ is used as the objective function value ($gc(D)$), then the feature selection problem is transformed into the following optimization problem $\max_S G(S)$.

It can be seen from the mathematical model that the feature selection is a discrete combinatorial optimization problem, so the MA algorithm can be used to solve this problem.

### 3.2. MA algorithm based on BDE and TS

MA is essentially a framework of optimization algorithm, which may use different search strategies to constitute different MA algorithms [21-23]. Considering the solution characteristics of the feature selection problem, BDE and TS are used as the global and local search tools in this paper.

#### 3.2.1 BDE-based global search

BDE proposed by Tao Gong extends differential evolution algorithm by modifying mutation operation to solve discrete optimization problems, which has the advantages of strong global optimization ability [29]. Here, BDE is used for global search to the entire solution space with the purpose of locating the optimal solution and providing good initial solution for local searching.

Appropriate control parameters have an important influence on performance of the BDE algorithm, so an adaptive tuning strategy of control parameters is adopted here. In the optimization process, the group fitness variance $\sigma^2$ reflects the population aggregation degree, which is suitable for dynamic adjustment of the control parameters. The $\sigma^2$ is defined as

$$\sigma^2 = \sum_{i=1}^{N_p} \left( \frac{f_i - f_{avg}}{f_{best}} \right)^2 \tag{6}$$

Where, $f_i$ is the fitness value of the $i$th individual, $f_{avg}$ is the average fitness value of the current population, $f_{best}$ is the best fitness for the population, $N_p$ is the population size.

Considering the distribution of population, the adaptive tuning strategy of the control parameters in BDE algorithm is given by:

$$F^g = F_{max} - (F_{max} - F_{min})(1 - \sigma_g^2 / N_p) \tag{7}$$

$$C_R^g = C_{R\min} + (C_{R\max} - C_{R\min})(1 - \sigma_g^2/N_p) \qquad (8)$$

Where, $F_{\max}$, $F_{\min}$ are respectively the maximum and minimum of the scale factor $F$, $C_{R\max}$ and $C_{R\min}$ are the maximum and minimum of the cross factor $C$, $g$ is the current evolution generation, $\sigma_g^2$ is the group fitness variance of the $g$th algebra. It can be seen that the scaling factor $F$ gradually decreases and cross-factor $C_R$ increases, when $\sigma^2$ is gradually decreased with the ongoing evolution of the population.

### 3.2.2. TS-based local search

Local search is an important concept in MA, which has an important impact on the algorithm's convergence speed and accuracy. The basic idea of local search is to search for better solutions constantly in the neighborhood of the current solution based on greedy thoughts. In this paper the neighborhood function is two switching operations 2-opt algorithm. TS is a heuristic optimization technique [30], which searches the solution space using the recent searching memory and contempt guidelines. TS has the following advantages: memory function, high search efficiency and strong climbing ability. However, the performance of TS is strongly dependent on initial solution.

TS is used as a local search strategy to deep intensification with the elite individual obtained by global search in each generation through BDE, i.e. search the local optimal solutions through continuous iteration in the local areas that the optimal solutions may exist. This 'coarse and fine' search strategy can comprehensively utilize the advantages of both BDE and TS, and make a balance between the solution quality and convergence speed.

### 3.3. Realization of feature selection

#### 3.3.1. Coding scheme

Considering the solution characteristics of the feature selection problem, a binary code coding scheme is used. The solution is encoded as a binary string, and the string length $CL$ is the number of all the original features ($CL=33$, here). In the binary string, each digit with value "1" or "0" represents whether the feature corresponding to the digit is selected.

#### 3.3.2. Determination of the fitness function

An appropriate fitness function is very important in MA algorithm, since it is the basis for guiding the search direction in the optimization process. In this paper, the fitness function is a KFRS-based generalized classification function, $Fitness = gc(D)$.

#### 3.3.3. The algorithm processes

The flow chart of the MA-based feature selection method is shown in Fig. 1. The specific steps of the process are described as follows.

Step 1: Data pre-processing. The z-score standardization method is used as the data pre-processing method for the obtained sample set.

Step 2: Parameter initialization. In BDE, the population size $N_p$ is 80, the maximum evolution generation is 300, the control factor $F$ and $C_R$ values ranged as [0.4, 0.9] and [0.3, 0.8]. In TS, the length of Tabu list $TL=20$, iterative steps is 200.

Step 3: Population initialization. Randomly generate $N_p$ solutions in the value range of control variables, obtain the new initial population by replacing the original individuals with better alternatives in the neighborhoods (5 neighborhood solutions) of the original ones, and set $g=1$.

Step 4: Calculation of the individual fitness values according to the fitness function.

Step 5: Global search. According to optimization mechanism of BDE, constantly updates the individual position through the mutation, crossover and selection.

Step 6: Local search. For elite individuals obtained by the global search, local search is done by TS. If a better solution in the corresponding neighborhood of the elites is found, the optimal solution will be updated.

Step 7: Judgment of termination conditions. Here, the termination condition is the evolutionary generations exceeds the maximum generation or the fitness value is greater than 0.9950. The optimization process will end if the termination condition is met; otherwise, the evolutionary generation $g$ is increased to $g+1$ and go to Step 4.

Step 8: Output the optimal solution and the corresponding feature subset.

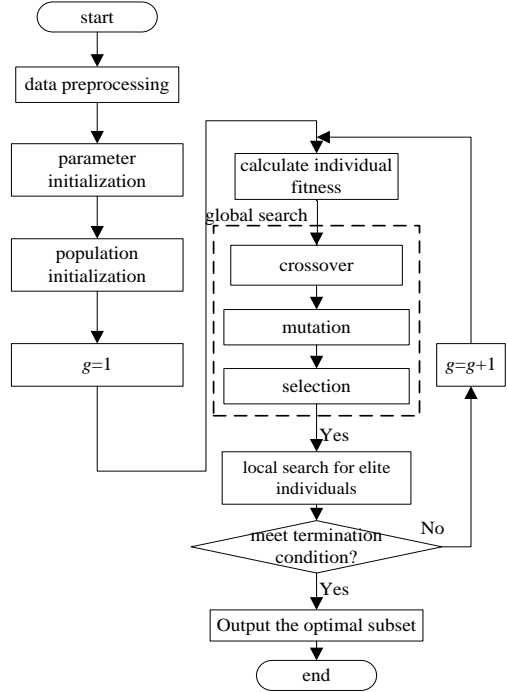

Fig. 1. Flowchart of feature selection based on MA

## 4. Construction of the original feature set

In this paper, the original feature set is constructed based on the following principles:

1) Mainstream principle—through deep analysis of the nature of power system transient stability, the selected features should have strong correlation with the system stability.

2) Real-time principle—based on the synchronized measurements provided by WAMS, the selected features should be able to characterize post-fault operation state of power systems.

3) Systematic principle—the selected features should be system-level features, rather than single-machine features, to ensure the number of the original input features not increasing with the increase of the system size, which will be suitable for stability analysis of large-scale power systems.

After having studied the existing literature comprehensively and carried out large amount of simulation, a group of system-level classification features whose number is independent of the scale of power systems are selected as the original features, as listed in Table 1. Here, $t_0$ and $t_{cl}$ denote the fault occurrence and clearing time in turn, $t_{cl+3c}$, $t_{cl+6c}$ and $t_{cl+9c}$ respectively denotes the 3rd, 6th and 9th cycle after the fault clearance.

Table 1

The original input features

| No. | Input features |
|---|---|
| Tz1 | Mean value of all the mechanical power before the fault incipient time |
| Tz2 | Maximum value of all the initial rotor acceleration rates |
| Tz3 | Initial rotor angle of the machine with the maximum acceleration rate |
| Tz4 | Mean value of all the initial acceleration power |
| Tz5 | Value of system impact at $t_{cl}$ |
| Tz6 | Rotor angle of the machine with the biggest difference relative to the centre of inertia at $t_{cl}$ |
| Tz7 | Kinetic energy of the machine with the maximum rotor angle at $t_{cl}$ |
| Tz8 | Rotor angle of the machine with the maximum kinetic energy at $t_{cl}$ |
| Tz9 | Maximum value of all the rotor kinetic energies at $t_{cl}$ |
| Tz10 | Mean value of all the rotor kinetic energies at $t_{cl}$ |
| Tz11 | Maximum value of the difference of rotor angles at $t_{cl}$ |
| Tz12 | Rotor angular velocity of the machine with the biggest difference relative to the centre of inertia at $t_{cl}$ |
| Tz13 | Value of system impact at $t_{cl+3c}$ |
| Tz14 | Maximum value of all the rotor kinetic energies at $t_{cl+3c}$ |
| Tz15 | Mean value of all the rotor kinetic energies at $t_{cl+3c}$ |
| Tz16 | Rotor angle of the machine with the biggest difference relative to the centre of inertia at $t_{cl+3c}$ |
| Tz17 | Maximum value of the difference of rotor angles at $t_{cl+3c}$ |
| Tz18 | Kinetic energy of the machine with the maximum rotor angle at $t_{cl+3c}$ |
| Tz19 | Rotor angular velocity of the machine with the biggest difference relative to the centre of inertia at $t_{cl+3c}$ |
| Tz20 | Value of system impact at $t_{cl+6c}$ |
| Tz21 | Maximum value of all the rotor kinetic energies at $t_{cl+6c}$ |
| Tz22 | Mean value of all the rotor kinetic energies at $t_{cl+6c}$ |
| Tz23 | Kinetic energy of the machine with the maximum rotor angle at $t_{cl+6c}$ |
| Tz24 | Rotor angle of the machine with the biggest difference relative to the centre of inertia at $t_{cl+6c}$ |
| Tz25 | Maximum value of the difference of rotor angles at $t_{cl+6c}$ |
| Tz26 | Rotor angular velocity of the machine with the biggest difference relative to the centre of inertia at $t_{cl+6c}$ |
| Tz27 | Value of system impact at $t_{cl+9c}$ |
| Tz28 | Kinetic energy of the machine with the maximum rotor angle at $t_{cl+9c}$ |
| Tz29 | Maximum value of all the rotor kinetic energies at $t_{cl+9c}$ |
| Tz30 | Mean value of all the rotor kinetic energies at $t_{cl+9c}$ |
| Tz31 | Rotor angle of the machine with the biggest difference relative to the centre of inertia at $t_{cl+9c}$ |
| Tz32 | Maximum value of the difference of rotor angles at $t_{cl+9c}$ |
| Tz33 | Rotor angular velocity of the machine with the biggest difference relative to the centre of inertia at $t_{cl+9c}$ |

## 5. Case study

The effectiveness of the proposed method is tested by the New England 39-bus power system and the southern power system of Hebei province. All of the programs are implemented in MATLAB on a PC platform with the master frequency 1.81 GHz and main memory 1 GB.

### 5.1. Case 1—the New England 39-bus power system

The New England 39-bus power system is a well-known test system for TSA studies reported in the literature [10-13, 15-17]. The one-line diagram of the power system is shown in Fig. 2.

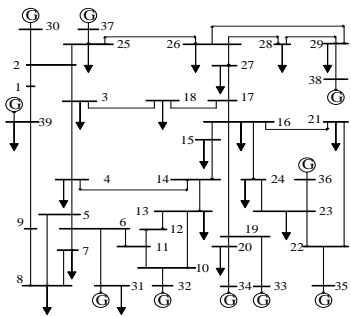

Fig. 2. New England 39-bus test system

#### 5.1.1. Generation of the sample sets

Extensive time domain simulation work has been carried out to create the training and test sample sets. The simulation is done with the four-order machine model and the IEEE DC1 excitation system model, as well as the constant impedance load model. A three-phase short-circuit faults is created at instant 0 s and cleared at 0.1 s. A successful reclosure of the faulted line is applied after fault clearance with no topology change from the fault. A total of 1100 arbitrary samples at 20 different fault locations is created under 80%, 85%, ……, 130% of the basic load levels. Corresponding to each loading level, 5 different generator outputs are randomly set. A total of 726 samples are randomly selected as the training data, and the remains as the testing data.

A class label "-1" or "+1" is assigned to each sample according to maximum relative rotor angle deviation during the transient period. If the maximum relative rotor angle deviation exceeded 360 degree [12], the class label is given as "-1" and the system is considered to be transiently unstable; otherwise, the label is given as "+1" and the system is stable.

### 5.1.2. Results of feature selection

Using the proposed feature selection approach, an optimal feature subset {Tz1, Tz4, Tz19, Tz24, Tz26, Tz31, Tz33} (named as $A_1$) is extracted from the original feature set. At the same time, a comparative test is carried out using some other algorithms, such as genetic algorithm (GA), discrete particle swarm optimization (DPSO) and BDE. The common parameters of the MA algorithm and the three other algorithms, such as population size and maximum number of iterations, are given the same values, and the other parameters are set as follows: crossover probability and mutation probability 0.85 and 0.01 respectively in GA, and the learning factor $c_1 = c_2 = 2$, inertia weight $\omega_0=1$ in DPSO. Considering the randomness of intelligent optimization algorithms, all the algorithms are independently run 100 times, and the results are shown in Table 2, where the search time is the average search time of 100 times.

Table 2

Training results of different optimization algorithms

| Optimization algorithm | Search time /s | Optimal fitness value | Search success rate /% |
|---|---|---|---|
| GA | 250.62 | 0.9571 | 79 |
| DPSO | 132.86 | 0.9487 | 53 |
| BDE | 78.53 | 0.9629 | 61 |
| MA | 83.79 | 0.9988 | 95 |

From Table 2, it can be seen that all the optimization algorithms can effectively complete the feature selection task. Since the optimization mechanisms of the algorithms are different, the proposed approach and BDE have better performance than the other two algorithms, such as higher fitness value, shorter search time and higher search success rate. At the same time, MA adopts an adaptive adjustment mechanism for control parameters and the 'coarse and fine' search strategy, thus it has the best result and the most stable performance.

In the optimization process, the fitness value evolution curves for the four different optimization algorithms are shown in Fig. 3.

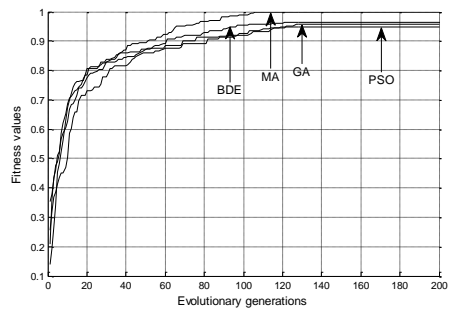

Fig. 3. Fitness curves of different optimization algorithms

Fig. 3 shows that all the algorithms are obviously effective for feature selection of PRTSA. In the four algorithms, MA has the highest convergence speed, which achieves the optimal result at 107th iteration, and its fitness value is higher than the others. Therefore, the proposed approach can improve the solution quality and search efficiency by comprehensively utilizing the advantages of BDE and TS.

### 5.1.3. Classification Test of the optimal feature subsets

In order to test the classification performance of the optimal feature subsets obtained, the feature subset $A_1$ obtained by the proposed method is used as the input of the commonly used classifier models of TSA, including multilayer perception (MLP)[15,16], SVM[7,12], and ELM[9]. Meanwhile, a comparative test is carried out for the feature subset $A_1$, the original feature set A, and a 10-dimensional feature subset $A_2$ obtained by PCA (with 95% of the variance of A). The parameters of the classifier models are set as follows. The MLP is designed with the hidden neuron number 15, and the back-propagation algorithm with the learning rate and momentum factor 0.8 and 0.7 respectively is employed. The kernel function of SVM used is RBF kernel and the associated parameters are optimized through a grid search during the 5-fold cross-validation process. For the ELM classifier, the hidden layer node number is set to 50.

Considering that the test accuracy $a$ has some kind of occasionality, the test results were assessed in statistical basis. The Kappa statistic value $K$ and the area under the receiver operating characteristic curve $r$ is used to assess the classification performance Where, $K$ is used to measure the consistency between prediction result and the actual classification, $r$ is the commonly used indicator to evaluate the overall performance of a classifier. If a classifier model is perfect, $r$ will be 1. If the model is just a random guess model, $r$ will be 0.5. The value of model of $r$ is greater, the model is more effective. Considering the above three classification performance indicators, a composite indicator $\eta$ is used to comprehensively evaluate the TSA classifier models' performance. $\eta$ is defined as

$$\eta = \frac{a + K + r}{3} \tag{9}$$

The test results of stability classification are given in Table 3.

Table 3

Test results on the New England 39-bus power system

| TSA model | Feature set | Dimension | $a$ /% | $K$ | $r$ | $\eta$ |
|---|---|---|---|---|---|---|
| MLP | A | 33 | 96.26 | 0.924 | 0.9666 | 0.9511 |
| | $A_1$ | 7 | 95.99 | 0.919 | 0.9660 | 0.9483 |
| | $A_2$ | 10 | 95.19 | 0.903 | 0.9495 | 0.9348 |
| SVM | A | 33 | 98.13 | 0.962 | 0.9832 | 0.9755 |
| | $A_1$ | 7 | 97.86 | 0.957 | 0.9817 | 0.9724 |
| | $A_2$ | 10 | 97.06 | 0.940 | 0.9744 | 0.9617 |
| ELM | A | 33 | 98.40 | 0.968 | 0.9829 | 0.9783 |
| | $A_1$ | 7 | 98.66 | 0.973 | 0.9869 | 0.9822 |
| | $A_2$ | 10 | 97.59 | 0.951 | 0.9724 | 0.9664 |

As shown in Table 3, compared with the original feature set A, $A_1$ has similar classification accuracy, but the data dimension is reduced to 7 from 33. Meanwhile, although the proposed method selects the less features than PCA, its classification performance is better than the latter. At the same time, it can be seen that the three TSA models have similar classification performance using $A_1$ and A, therefore the optimal feature subset obtained has a good suitability for different TSA models.

### 5.2. Case 2 – the southern power system of Hebei province

The southern power system of Hebei province is next employed to demonstrate the effectiveness of the proposed approach of feature selection for a more complex and practical system.

### 5.2.1. Generation of the sample sets

Extensive simulation has been carried out to generate the sample sets. Of all 83 generators in the system, 11 generators are modeled as the six-order model, and the excitation systems and governors are considered; others the classical machine model. The load model is represented by a comprehensive model with 40% constant-impedance and 60% constant power.

In the range from 90% to 120% of the basic load level, active and reactive powers of generators are set correspondingly. A three-phase short-circuit fault is created at instant 0.1 s and cleared at 0.2 s. A successful reclosure of the faulted line is applied after fault clearance with no topology change from the fault. The fault locations lie at 0, 25%, 50%, and 75% of the length on transmission lines. The stability criterion is as same as in Case-1. A total of 2000 samples are generated, 1320 of which are randomly selected to build up the training set, and the remains the test set.

### 5.2.2. Classification Test of the optimal feature subsets

The optimal feature subset obtained by the proposed method is {Tz4, Tz9, Tz16, Tz17, Tz18, Tz24, Tz25, Tz26, Tz31, Tz32, Tz33} (named as $B_1$). A 13-dimensional feature subset $B_2$ is given by the PCA method. A comparative test is carried out for the original feature set A, $B_1$, $B_2$ and $A_1$ (the optimal subset in Case-1) with the results shown in Table 4.

Table 4

Test results on the southern power system of Hebei province

| TSA model | Feature set | Dimension | $a$ /% | $K$ | $r$ | $\eta$ |
|---|---|---|---|---|---|---|
| MLP | A | 33 | 92.65 | 0.844 | 0.9163 | 0.8956 |
| | $B_1$ | 11 | 92.35 | 0.840 | 0.9146 | 0.8927 |
| | $B_2$ | 13 | 91.76 | 0.828 | 0.9134 | 0.8863 |
| | $A_1$ | 7 | 91.18 | 0.813 | 0.8950 | 0.8733 |
| SVM | A | 33 | 95.74 | 0.910 | 0.9414 | 0.9363 |
| | $B_1$ | 11 | 95.44 | 0.903 | 0.9511 | 0.9362 |
| | $B_2$ | 13 | 95.00 | 0.895 | 0.9321 | 0.9257 |
| | $A_1$ | 7 | 94.12 | 0.876 | 0.9256 | 0.9143 |
| ELM | A | 33 | 95.88 | 0.913 | 0.9576 | 0.9431 |
| | $B_1$ | 11 | 96.47 | 0.925 | 0.9617 | 0.9505 |
| | $B_2$ | 13 | 95.59 | 0.907 | 0.9508 | 0.9379 |
| | $A_1$ | 7 | 94.85 | 0.892 | 0.9442 | 0.9282 |

From Table 4, it can be observed that $B_1$ has similar classification performances with A, but the data dimension is reduced to 11 from 33. The classification performance of $B_1$ is better than $B_2$. Furthermore, the composite indicator $\eta$ of $B_1$ is higher around 2.0% than $A_1$. This shows that as the size of the systems increases, complexity of the transient stability pattern increases, and the number of selected input features should also be increased accordingly to provide more adequate knowledge to fully reflect the power system stability characteristics.

### 6. Conclusions

Considering the possible real-time information provided by WAMS, a new feature selection method for transient stability assessment of power systems using KFRS and MA is presented in this paper. The proposed method is examined on the New England 39-bus test system and the southern power system of Hebei province. The following conclusions can be drawn from the work:

(1) MA algorithm can effectively solve the feature selection problem of PRTSA, and has better solution quality, shorter search time and higher search success rate than other optimization algorithms such as GA, DPSO and BDE.

(2) Without sacrificing the classification performance, the proposed method can significantly reduce the dimension of the original feature set and is superior to the commonly used PCA method. The obtained feature subsets can be generally applied to a variety of TSA classifier models, such as MLP, SVM and ELM, with satisfactory classification performance.

(3) The proposed approach may find potential applications in real-time transient stability prediction and online dynamic security assessment of power systems. Furthermore, the methodology of feature selection may be applied to any similar pattern classification problem in engineering field.